\documentclass[12pt,a4paper,twoside]{article}

\usepackage{fancyhdr,graphicx}
\usepackage{amsmath,amsfonts,amsthm}
\usepackage{psfrag}

\DeclareMathOperator{\tr}{Tr} 
\DeclareMathOperator{\rpart}{Re} 
\DeclareMathOperator{\ipart}{Im} 
\DeclareMathOperator{\diag}{diag}
\DeclareMathOperator{\supp}{Supp}
\DeclareMathOperator{\resi}{Res}

\numberwithin{equation}{section}

\newcommand{\res}{\resi\displaylimits}
\newcommand{\Lie}{\mathcal{L}}
\newcommand{\p}{\partial}

\newcommand{\abs}[1]{\left \lvert #1 \right \rvert}

\theoremstyle{plain}
\newtheorem{theorem}{Theorem}[section]
\newtheorem{proposition}{Proposition}[section]
\newtheorem{lemma}{Lemma}[section]
\newtheorem{corollary}{Corollary}[section]

\theoremstyle{remark}
\newtheorem{remark}{Remark}

\theoremstyle{definition}

\begin{document} 
\title{On an average over the Gaussian Unitary Ensemble
\footnote{The authors acknowledge financial
    support by the EPSRC grant EP/D505534/1.}}  
   \author{F. Mezzadri and M. Y. Mo}  
\date{}
\maketitle

\begin{abstract}
  We study the asymptotic limit for large matrix dimension $N$ of the
  partition function of the unitary ensemble ($\beta = 2$) with weight
\begin{equation*}
  w(x) := \exp\left(-\frac{z^2}{2x^2}+\frac{t}{x} -\frac{x^2}{2} \right).
\end{equation*}
We compute the leading order term of the partition function and of the
coefficients of its Taylor expansion.  Our results are valid in the
region $c_1N^{-\frac12} < z < c_2N^\frac14$.
Such partition function contains all the information on a new
statistics of the eigenvalues of matrices in the Gaussian Unitary
Ensemble (GUE) that was introduced by Berry and Shukla~\cite{BeSh}. It
can also be interpreted as the moment generating function of the
singular linear statistics
\begin{equation*}
  \sum_{j=1}^N\left(\frac{1}{x_j} + \frac{1}{x_j^2}\right).
\end{equation*}

\end{abstract}

\vspace{.25cm}

\hspace{.26cm} 2000 MSC: 15A52, 35Q15.

\section{Introduction}
\label{introduction}
\subsection{Background}
\label{background}
In Random Matrix Theory partition functions of ensembles whose
probability measure is invariant under conjugation by unitary matrices
(unitary or $\beta =2$ ensembles) are integrals of the form
\begin{equation}
  \label{eq:partition_functions}
 \frac{1}{N!} \int_{J^N} \prod_{j=1}^N w(x_j)\prod_{1 \le j < k \le N}\abs{x_k -
    x_j}^2 d^Nx,
\end{equation}
where $w(x)\ge 0$ is a weight function and usually $J$ is either an
interval, or the whole real line or the unit circle.  The theory of
orthogonal polynomials (see Szeg\H{o}~\cite{Sze75}, pp. 23--28)
implies that such integrals are determinants of Hankel or Toeplitz
matrices, or a linear combination of the two.  Computing asymptotic
formulae of such determinants is a very important task --- often a
very difficult one --- in many branches of mathematics and
physics. The asymptotics of the
integral~\eqref{eq:partition_functions} depends crucially on the
analytic property of $w(x)$. Usually, singular weights are the most
challenging.

The purpose of this article is to compute the following expectation
value over the Gaussian Unitary Ensemble (GUE):
\begin{equation}
  \label{eq:main_average}
  E_N(z,t) := \int_{\mathbb{R}^N}
  \Biggl( \prod_{j=1}^N \exp 
    \left(-\frac{z^2}{2x_j^2} + \frac{t}{x_j} \right)
  \Biggr)P_{\mathrm{GUE}}(x_1,\dotsc,x_N)
  d^Nx,
\end{equation}
where
\begin{equation}
  P_{\mathrm{GUE}}(x_1,\dotsc,x_N) :=
  \frac{1}{Z_N N!}\exp\left(-\frac{1}{2}\sum_{j=1}^Nx_j^2\right)\prod_{1\le j
    < k \le N}\abs{x_k - x_j}^2
\end{equation}
is the joint probability density function (\textit{j.p.d.f.}) of the
eigenvalues and
\begin{equation}
  \begin{split}
  Z_N & := \frac{1}{N!}\int_{\mathbb{R}^N}
    \exp\left(-\frac{1}{2}\sum_{j=1}^Nx_j^2\right)\prod_{1\le j
    < k \le N}\abs{x_k - x_j}^2 d^Nx \\
     & = \left(2\pi\right)^{N/2} \prod_{j=1}^{N-1} j!
  \end{split}
\end{equation}
is the partition function of the GUE.

Let 
\begin{equation}
  \Lambda_N(x) := \prod_{j=1}^N\left(x - x_j\right),
\end{equation}
be a polynomial of degree $N$ whose roots $x_1,\dotsc,x_N$ are all
real and define
\begin{equation}
  \label{eq:tuck_function}
  Q_N(x) := \frac{\Lambda_N'^2(x)}{\Lambda_N'^2(x) -\Lambda_N(x)
  \Lambda_N''(x)}.
\end{equation}
This function was studied by Tuck~\cite{Tuc08} in a numerical
investigation of the zeros of the Riemann zeta function $\zeta(s)$. In
Tuck's article $\Lambda_N(x)$ was replaced by the Hardy function:
\begin{equation}
  Z(t) := t^{1/4}\exp\left(\tfrac{\pi t}{4}\right)\pi^{-\tfrac12 i
    t}\Gamma\left(\tfrac14 + \tfrac12 i t\right)\zeta\left(\tfrac12 +
    it\right).
\end{equation}
The Hardy function is an entire function of order one; it is real for
$t \in \mathbb{R}$ and its zeros coincide with non trivial zeros of
$\zeta(s)$. The motivation to study~\eqref{eq:tuck_function} was that
the Riemann hypothesis implies that
\begin{equation}
   W(t) := Z'^2(t) -Z(t)Z''(t) >0. 
\end{equation}
More generally, if $Z(t)$ is an entire function of order less or equal
to one, real on $\mathbb{R}$ and whose zeros are all real then $W(t)>0$.

Suppose that the roots of $\Lambda_N(x)$ are $N$ random variables with
\textit{j.p.d.f.} $p(x_1,\dotsc,x_N)$, and denote by $P(Q_N)$ the
probability density function (\textit{p.d.f.}) of the random variable
$Q_N(x)$.  Berry and Shukla~\cite{BeSh} observed that $Q_N(x)$ is a
sensitive indicator of the degree of repulsion between neighbouring
zeros of $\Lambda_N(x)$. More precisely, the rate of decay of $P(Q_N)$
as $Q_N \to \infty$ is related to the rigidity of the roots of
$\Lambda_N(x)$.  They also studied in detail $P(Q_N)$ in two cases:
when the roots of $\Lambda_N(x)$ are $N$ independent identically
distributed (\textit{i.i.d.}) standard normal random variables and
when $\Lambda_N(x)$ is the characteristic polynomial of a matrix in
the GUE.

Berry and Shukla showed that all the information on $P(Q_N)$ is
contained in the expectation value
\begin{equation}
  \label{eq:general_average}
\int_{\mathbb{R}^N}\Biggl( \prod_{j=1}^N \exp 
    \left(-\frac{z^2}{2x_j^2} + \frac{t}{x_j} \right)
  \Biggr)p(x_1,\dotsc,x_N)d^Nx,
\end{equation}
in the sense that all the moments of the distribution $P(Q_N)$ can be
extracted from its knowledge. They computed the
average~\eqref{eq:general_average} and the moments of $P(Q_N)$ when
the roots of $\Lambda_N(x)$ are \textit{i.i.d.} standard normal random
variables, but not when they are the eigenvalues of matrix in the GUE.

The integral in  the right-hand side of
equation~\eqref{eq:main_average} is amenable to other
interpretations. For example, if we set
\begin{equation}
   s=z^2/2 = -t, \quad s>0, 
\end{equation}
then
\begin{equation}
  M_N(s) := \int_{\mathbb{R}^N}\prod_{j=1}^N \exp 
  \Biggl(-s \left(\frac{1}{x_j^2} + \frac{1}{x_j}\right) \Biggr)
   P_{\mathrm{GUE}}(x_1,\dotsc,x_N)d^Nx
\end{equation}
is the moment generating function of the \textit{p.d.f.} of the singular
linear statistics
\begin{equation}
  \sum_{j=1}^N \left(\frac{1}{x_j^2} + \frac{1}{x_j}\right).
\end{equation}
Furthermore, $Z_N E_N(z,t)$ is the partition function of the unitary
ensemble with weight
\begin{equation}
 \label{eq:new_weight}
  w(x) := \exp\left(-\frac{z^2}{2x^2}+\frac{t}{x} -\frac{x^2}{2} \right).
\end{equation}
Now, denote by $\pi_j(x)$, $j\in \mathbb{Z}_+$, the monic polynomials
orthogonal with respect to $w(x)$.  The integral $E_N(z,t)$
can be rewritten as a Hankel determinant:
\begin{equation}
   E_N(z,t)= Z_N^{-1}\det\left(\mu_{j + k}\right)_{j,k=0}^{N-1} 
  = Z_N^{-1}\prod_{j=0}^{N-1} h_j, 
\end{equation}
where
\begin{equation}
   \mu_j := \int_{-\infty}^\infty w(x) x^j dx, \quad j\in \mathbb{Z}_+
\end{equation}
and
\begin{equation}
  \int_{-\infty}^\infty w(x) 
    \pi_j(x)\pi_k(x)dx = h_j \delta_{jk}.
\end{equation}   
If we know the behaviour of the polynomials $\pi_N(x)$ as $N \to
\infty$ (Plancherel-Rotach asymptotics), then we can extract
information on the asymptotic limit of $E_N(z,t)$. Our approach
consists in studying the solution of the Riemann-Hilbert (R-H) problem
associated to the polynomials $\pi_N(x)$ in the limit as $N \to
\infty$. The main tool is the nonlinear steepest descent analysis
developed by Deift \textit{et al}~\cite{DKV,DKV2}.  The average
$E_N(z,t)$ can then be computed in terms of such a solution using a set
of differential identities introduced by Bertola \textit{et
  al}~\cite{BEH}.

After this work was completed, we discovered that independently Chen
and Its~\cite{CI08,CI09} studied the orthogonal polynomial problem
associated with the partition function
\begin{equation}
  \label{eq:CI_partition_fun}
    H_N(\alpha,s) :=\frac{1}{N!}\int_{[0,\infty)^N}
  \prod_{j=1}^Nx_j^\alpha e^{-x_j -s/x_j}\prod_{1\le j
    < k \le N}\abs{x_k - x_j}^2 d^Nx, 
\end{equation}
where $\alpha > - 1$ and $s \ge 0$. The weight in this integral is
that one of the Laguerre polynomials perturbed by the singular factor
$e^{-s/x}$.  In a first paper Chen and Its~\cite{CI08} proved that
$H_N(\alpha,s)$ can be expressed as the integral of the combination of
particular third Painlev\'e functions; in a second article~\cite{CI09}
they derived asymptotic formulae for the orthogonal polynomials, the
corresponding recurrence coefficients and the $h_j$'s. When the
parameters $t$ and $\alpha$ in equation~\eqref{eq:new_weight} and
$x^\alpha e^{-x -s/x}$ respectively are both set equal to zero, then
the systems of orthogonal polynomials associated with the two weights
can be mapped into each other by the change of variables $x \mapsto
x^2/2$.  The Plancherel-Rotach asymptotics of the orthogonal
polynomials can be studied using the nonlinear steepest descent in
both cases. However, even when $t=\alpha=0$, the partition
functions~\eqref{eq:main_average} and~\eqref{eq:CI_partition_fun} are
not equivalent and cannot be mapped into each other by a simple change
of variables.

\subsection{Statement of results}
The average~\eqref{eq:main_average} is an entire function of $t$,
thus its Taylor series has an infinite radius of convergence and
we can write
\begin{equation}
  \label{eq:E_series}
  E_N(z,t) = \sum_{m=0}^\infty E_{Nm}(z)t^m.
\end{equation}
The main goal of this paper is to compute the leading order term in the
 asymptotic expansion of $E_N(z,t)$ and  $E_{N  m}(z)$ as $N \to \infty$.
Our main result is the following:
\begin{theorem}
\label{thm:main}
Let $c_1N^{-\frac{1}{2}}<z<c_2N^{\frac14}$, where $c_1$ and $c_2$ are
two constants independent of $z$ and $N$.  The expectation
value $E_{N}(z,t)$ is given by
\begin{equation}
\label{eq:asymdet}
\begin{split}
  E_{N}(z,t)&=B_{N}\exp\left(\frac{z^2}{4}-
    \frac{9}{2^{\frac{10}{3}}}\left(N^{\frac{2}{3}}z^{\frac{4}{3}}-1\right)
    +\frac{t^2N^{\frac{1}{3}}}{2^{\frac{5}{3}}z^{\frac{4}{3}}}\right)
  \\
  &\quad  \times\bigl(1+o(1)\bigr), \quad N \to \infty, 
\end{split}
\end{equation}
where $B_{N}$ is the ensemble average
\begin{equation}
B_{N}:= \int_{\mathbb{R}^N}\left(\prod_{j=1}^N 
e^{-\frac{1}{2Nx_j^2}}\right) P_{\mathrm{GUE}}(x_1,\dotsc,x_N) d^Nx,
\end{equation}
which is independent of $t$.
\end{theorem}

Berry and Shukla~\cite{BeSh} showed that the \textit{m}-th moment of
$P(Q_N)$ is given by the integral
\begin{equation}
\label{eq:moments}
M_{Nm}: =2^{1-m}\prod_{n=m}^{2m}n\int_{0}^{\infty}z^{2m-1}E_{N 2m}(z)dz.
\end{equation}
From Theorem~\ref{thm:main} it is straightforward to compute the
coefficient $E_{N 2m}(z)$ in the series
expansion~\eqref{eq:E_series}.
\begin{corollary}
\label{cor:main}
Let $c_1N^{-\frac{1}{2}}<z<c_2N^{\frac14}$, where $c_1$ and $c_2$ are
two constants independent of $z$ and $N$. The leading order term of
the coefficient of $t^{2m}$ in equation~\emph{(\ref{eq:E_series})} is
\begin{equation}
  \label{eq:main}
    E_{N 2m}(z) \sim B_{N} \exp\left(\frac{z^2}{4}
-\frac{9}{2^{\frac{10}{3}}}\left(N^{\frac{2}{3}}z^{\frac{4}{3}}-1\right)\right)
\frac{N^{\frac{m}{3}}}{2^{\frac{5m}{3}}m!z^{\frac{4m}{3}}}, \quad N
\to \infty.
\end{equation}
\end{corollary}

Unfortunately, the asymptotic limit in equations~\eqref{eq:asymdet}
and~\eqref{eq:main} cannot be assumed to be uniform in $z$: there may
be non negligible contributions from the region $z < c_1
N^{-\frac{1}{2}}$ that would affect the
integral~\eqref{eq:moments}. Such contributions can be investigated by
studying the double scaling limit of the matrix ensemble with
weight~\eqref{eq:new_weight}. This will be the subject of a
forthcoming publication.

The structure of this article is the following: in
\S\ref{preliminaries} we introduce the R-H problem for the orthogonal
polynomials with weight~\eqref{eq:new_weight} (after appropriate
rescaling of $x$, $z$ and $t$) and the differential identities used to
compute the leading order asymptotics of $E_N(z,t)$; in
\S\ref{se:equilibrium_measure} we find the equilibrium measure on
which the R-H analysis is based; in \S\ref{se:RH} we apply the
nonlinear steepest descent to the R-H problem; \S\ref{se:asy_diffid}
and \S\ref{sec:ensemble} are devoted to complete the proof of
Theorem~\ref{thm:main} combining the asymptotics of the solution of
the R-H problem and the differential identities discussed in
\S\ref{preliminaries}.

\subsection*{Acknowledgements}
We would like to thank Professor Sir Michael Berry for introducing us to the
problem tackled in this paper.  We are also indebted for helpful
discussions to Professor Alexander Its and Dr Igor Krasovsky.  We are
particularly grateful to Professor Alexander Its for making available
to us the manuscript~\cite{CI09} before publication.

\section{Preliminaries}
\label{preliminaries}
Let us introduce the scaling
\begin{equation}
 v_1 = \frac{t}{\sqrt{N}}, \quad v_2 = \left(\frac{z}{N}\right)^2 \quad
 \text{and} \quad y_j = \frac{x_j}{\sqrt N}, \quad j=1,\dotsc N.
\end{equation}
The  weight~\eqref{eq:new_weight} becomes
\begin{equation}
  w_N(y):=  \exp\Biggl(-N\left(\frac{v_2}{2y_j^2} + 
          \frac{y_j^2}{2}\right) +
    \frac{v_1}{y_j} \Biggr).
\end{equation}
We also define the partition function
\begin{equation}
  \label{eq:tildeB}
    G_N(v_1,v_2)  := \frac{1}{N!}\int_{\mathbb{R}^N}
     \prod_{j=1}^N\exp\Biggl(-N\left(\frac{v_2}{2y_j^2} +
       \frac{y_j^2}{2}\right)
          + \frac{v_1}{y_j} \Biggr)
         \prod_{1 \le j < k \le N}^N\abs{y_k-y_j}^2 d^Ny,
\end{equation}
which is proportional to the average $E_N(z,t)$, namely
\begin{equation}
  E_N(z,t) = Z_N^{-1}N^{\frac{N^2}{2}}G_N(v_1,v_2).
\end{equation}
For convenience, where there is no risk of confusion with the
quantities introduced in \S\ref{background}, we denote the
polynomials orthogonal with respect to $w_N(y)$ by $\pi_j(y)$.
Similarly, we write
\begin{equation}
  \label{eq:hankel_det_2}
  G_N(v_1,v_2) = \det\left(\mu_{j + k}\right)_{j,k=0}^{N-1} 
  = \prod_{j=0}^{N-1}h_j, 
\end{equation}
where
\begin{equation}
 \mu_j := \int_{-\infty}^\infty w_N(y) y^j dx, 
  \quad j\in \mathbb{Z}_+
\end{equation}
and
\begin{equation}
  \label{eq:orthogonality_2}
  \int_{-\infty}^\infty w_N(y) \pi_j(y)\pi_k(y)dy = h_j \delta_{jk}.
\end{equation}

Let us define the matrix valued function
\begin{equation}
   Y(y) := 
   \begin{pmatrix} 
   \pi_N(y) & \frac{1}{2\pi i}
    \int_{-\infty}^\infty\frac{\pi_N(s)w_N(s)}{s-y}ds  \\
     \kappa_{N-1}\pi_{N-1}(y) & 
     \frac{\kappa_{N-1}}{2\pi i}
      \int_{-\infty}^\infty\frac{\pi_{N-1}(s)w_N(s)}{s-y}ds
\end{pmatrix},
\end{equation}
where $\kappa_{N-1}=-2\pi ih_{N-1}$. Fokas \textit{et
  al}~\cite{FIK91,FIK92} showed that $Y(y)$ solves the following R-H
problem:
\begin{equation}
  \label{eq:RHP}
\begin{aligned}
1. \quad  & \text{$Y(y)$ is analytic in $\mathbb{C}/\mathbb{R}$},&&\\
2. \quad & Y_+(y)=Y_-(y)\begin{pmatrix} 1 & w_N(y)  \\ 0 & 1
\end{pmatrix},& & y\in\mathbb{R},\\
3. \quad  & Y(y)=\left(I+O(y^{-1})\right)\begin{pmatrix} y^N & 0
\\ 0 & y^{-N} \end{pmatrix}, &\quad  & y\rightarrow\infty
\end{aligned}
\end{equation}
where $Y_+(y)$ and $Y_-(y)$ denotes the limiting values of
$Y(y)$ as it approaches the left and right-hand side of the real
axis.  It turns out that the partition function $G_N(v_1,v_2)$ can
be expressed in terms of $Y(y)$.
\begin{lemma}[Bertola, Eynard and Hanard~\cite{BEH}]
\label{thm:beh}
The following differential identities hold:
\begin{subequations}
\label{eq:diffid}
\begin{align}
\label{eq:diffidv1}
\frac{\p \log G_N}{\p v_1}&=-\frac{1}{4\pi
i}\oint_{y=0}\frac{1}{y}\tr\left(Y^{-1}(y)Y^{\prime}(y)\sigma_3\right)dy,\\
\label{eq:diffidv2}
\frac{\p \log G_N}{\p v_2}&=\frac{N}{8\pi i}\oint_{y=0}\frac{1}{y^2}
\tr\left(Y^{-1}(y)Y^{\prime}(y)\sigma_3\right)dy,
\end{align}
\end{subequations}
where the contour of integration is a small loop around $y=0$ oriented
counter-clockwise.   
\end{lemma}
\begin{proof}
  Taking the logaritmic derivatives of both sides of
  equation~\eqref{eq:hankel_det_2} and using the orthogonality
  conditions~\eqref{eq:orthogonality_2} gives
\begin{equation}
\frac{\p\log G_N}{\p v_k}=\int_{-\infty}^\infty\left(\sum_{j=0}^{N-1}
\frac{\pi_j^2(y)}{h_j}\right)\frac{\p w_N(y)}{\p
  v_k}dy,\quad k=1,2.
\end{equation}
These integrals can be rewritten as
\begin{equation}
\label{eq:logder2}
\frac{\p\log G_N}{\p
v_k}=\left(-N\right)^{k-1}
\int_{-\infty}^\infty\frac{K_N(y,y)}{ky^k}dy,\quad k=1,2,
\end{equation}
where $K_N(x,y)$ is the kernel
\begin{equation}
K_N(x,y):=\sqrt{w_N(x)w_N(y)}
\sum_{j=0}^{N-1}\frac{\pi_j(x)\pi_j(y)}{h_j}
\end{equation}
and
\begin{equation}
  G_N(v_1,v_2) = \left(N!\prod_{j=0}^{N-1} h_j\right) \det
  \bigl(K_N(y_j,y_k)\bigr)_{j,k=0}^{N-1}.
\end{equation}
In order to evaluate the integral~\eqref{eq:logder2} we use the relation
\begin{equation}
\begin{split}
K_{N}(x,y)&=\frac{\sqrt{w_N(x)w_N(y)}}{2\pi
  i(x-y)}\begin{pmatrix} 0 & 1 \end{pmatrix}
  Y_+^{-1}(y)Y_+(x)\begin{pmatrix} 1 \\ 0 \end{pmatrix}\\
&=\frac{\sqrt{w_N(x)w_N(y)}}{2\pi
i(x-y)}\tr\Biggl(Y_+^{-1}(y)Y_+(x)\begin{pmatrix} 0 &1\\ 0&0
\end{pmatrix}\Biggr),
\end{split}
\end{equation}
which follows from the Christoffel-Darboux formula, the definition of
$Y(x)$ and the fact that $\det Y(y) =1$. By using l'Hospital's rule
we obtain
\begin{equation}
\label{eq:limitkern}
K_{N}(y,y)= \frac{w_N(y)}{2\pi i} 
\tr\Biggl(Y_+^{-1}(y)Y_+^{\prime}(y)\begin{pmatrix} 0 &1\\ 0&0
\end{pmatrix}\Biggr).
\end{equation}
Combining the jump condition of the R-H problem~\eqref{eq:RHP} and
equation \eqref{eq:limitkern} gives
\begin{equation}
K_N(y,y)=\frac{1}{4\pi
i}\Bigl(\tr\left(Y_-^{-1}(y)Y_-^{\prime}(y)\sigma_3\right)\bigr)
-\tr\left(Y_+^{-1}(y)Y_+^{\prime}(y)\sigma_3\right)\Bigr),\quad
y\in\mathbb{R}.
\end{equation}
The asymptotic behaviour of $Y(y)$ as $y \to \infty$ implies that the
functions
\begin{equation}
\tr\left(Y_{\pm}^{-1}(y)Y_{\pm}^{\prime}(y)\sigma_3\right)
\end{equation}
are analytic in the upper/lower half planes with a simple pole at
infinity. Finally, the identities~\eqref{eq:diffid} follow from the
residue theorem.
\end{proof}

This lemma gives an explicit link between the solution of the R-H
problem and the average~\eqref{eq:main_average}.  The main challenge
that we are facing is to compute an asymptotic formula for $Y(y)$
using the nonlinear steepest descent method. Then, we can obtain a
formula for $E_N(z,t)$ using equations~\eqref{eq:diffid}.

\section{The equilibrium measure}
\label{se:equilibrium_measure}
The asymptotics analysis of $Y(y)$ relies on computing the
\textit{g}-function and the support of the equilibrium
measure for the potential 
\begin{equation}
V_0(y) :=\frac{v_2}{2y^2} + \frac{y^2}{2},
\end{equation}
where we have neglected the term $v_1/y$ because it is asymptotically
small. The equilibrium measure is the positive normalized Borel
measure $\mu(y)$ that minimizes the energy function
\begin{equation}
I(\mu) :=-\int^\infty_{-\infty}\int^\infty_{-\infty}
\log|x-y|d\mu(x)d\mu(y)+\int^\infty_{-\infty}V_0(y)d\mu(y).
\end{equation}
It satisfies the conditions
\begin{subequations}
\label{eq:ineq}
\begin{alignat}{2}
&2\int_{-\infty}^\infty\log |y-s|d\mu(s)-V_0(y)= l, & &
x\in\supp \left (\mu \right ),\\
&2\int_{-\infty}^\infty\log|y-s|d\mu(s)-V_0(y)\leq l,  &\qquad &
x\in\mathbb{R}/\supp\left(\mu\right).
\end{alignat}
\end{subequations}
for some constant $l$. Moreover, if $\mu(y)$ satisfies
\eqref{eq:ineq}, then it must be equal to the equilibrium measure.

The \textit{g}-function is defined by
\begin{equation}
g(y) :=\int^\infty_{-\infty}\log (y-s)d\mu(s).
\end{equation}
It is analytic in $\mathbb{C}\setminus\mathbb{R}$ and has jump
discontinuities on the real axis. The conditions \eqref{eq:ineq}
expressed in terms of $g(y)$ become
\begin{subequations}
\label{eq:ineqg}
\begin{alignat}{2}
&g_+(y)+g_-(y)-V_0(y)= l, 
&& y\in \supp\left(\mu\right),\\
&\mathrm{Re}\bigl(g_+(y)+g_-(y)\bigr)-V_0(y)\leq l, 
& \qquad &y\in\mathbb{R}/\supp\left(\mu\right).
\end{alignat}
\end{subequations}
Furthermore, we have
\begin{equation}
g(y)=\log y+O(y^{-1}), \quad  y \to \infty. 
\end{equation}

We shall derive an expression for $g(y)$ using the ansatz
\begin{equation}
g^{\prime}(y)=\frac{V_0^{\prime}(y)}{2}-
\frac{(y^2-\lambda_1^2)\sqrt{(y^2-\lambda_2^2)(y^2-\lambda_3^2)}}{2y^3}.
\end{equation}
The constants $\lambda_1$, $\lambda_2$ and $\lambda_3$ are
determined by the asymptotic behaviour of $g'(y)$, namely
\begin{subequations}
\begin{align}
g^{\prime}(y) & =\frac{1}{y}+O(y^{-2}), \qquad y \to \infty, \\
g'(y) & = O(1), \qquad   y \to 0.
\end{align}
\end{subequations}
These constraints give 
\begin{subequations}
\label{eq:aj}
\begin{align}
\label{eq:aj_1}
&\lambda_1^2+\frac{1}{2}\left(\lambda_2^2+\lambda_3^2\right)=2,\\
\label{eq:aj_2}
&\lambda_1^{-2}+\frac{1}{2}\left(\lambda_2^{-2}+\lambda_3^{-2}\right)=0,\\
\label{eq:aj_3}
&\lambda_1^2\lambda_2\lambda_3=-v_2.
\end{align}
\end{subequations}
Let $A_j=\lambda_j^2$. Then, we see that $A_2$, $A_3$ are the
solutions of the quadratic equation
\begin{equation}
\label{eq:A23}
y^2-2(2-A_1)y+\frac{v_2^2}{A_1^2}=0,
\end{equation}
while $A_2^{-1}$ and $A_3^{-1}$ are solutions of
\begin{equation}\label{eq:A23-}
y^2+\frac{2}{A_1}y+\frac{A_1^2}{v_2^2}=0.
\end{equation}
From equation \eqref{eq:A23} it follows that $A_2^{-1}$ and $A_3^{-1}$
also satisfy
\begin{equation}
\frac{v_2^2}{A_1^2}y^2-2(2-A_1)y+1=0.
\end{equation}
By comparing this equation with (\ref{eq:A23-}), we see that $A_1$ is a
solution of the equation
\begin{equation}\label{eq:A1}
A_1^4-2A_1^3-v_2^2=0.
\end{equation}
\begin{lemma}
\label{le:realA}
There exists a solution to equation~\textrm{\eqref{eq:aj}} such that
$\lambda_1 \in i \mathbb{R}$ and $\lambda_2, \lambda_3 \in
\mathbb{R}$.
\end{lemma}
\begin{proof} The solutions of (\ref{eq:A23-}) are 
\begin{equation}\label{eq:sol23}
y =-\frac{1}{A_1}\left(1\pm\sqrt{1-\left(\frac{A_1}{\sqrt{v_2}}\right)^4}\right)
\end{equation}
Therefore, if a solution of (\ref{eq:A1}) such that
$-\sqrt{v_2}<A_1<0$ exists, then the two solutions to (\ref{eq:sol23})
are real and positive and the lemma is proven. Now,  if $A_1=0$, then 
the left-hand side  of (\ref{eq:A1}) is $-v_2^2<0$;  if
$A_1=-\sqrt{v_2}$ it is $2v_2^{\frac{3}{2}}>0$. Hence, there
is a solution to (\ref{eq:A1}) between $-\sqrt{v_2}$ and $0$.  
\end{proof}

Let us now choose $\lambda_1$, $\lambda_2$ and $\lambda_3$ such that 
$\ipart(\lambda_1)>0$ and $0<\lambda_2<\lambda_3$. Furthermore, define 
\begin{equation}
\label{eq:gfun}
g(y) :=\frac{V_0(y)}{2}-\int_{\lambda_3}^y\frac{(s^2-\lambda_1^2)
\sqrt{(s^2-\lambda_2^2)(s^2-\lambda_3^2)}}{2s^3}ds+\frac{l}{2},
\end{equation}
where the integration path is chosen such that it does not intersect
the interval $(-\infty,\lambda_3)$ and 
\begin{equation}
\label{eq:ell}
l=-2\lim_{y\rightarrow\infty}\left(\frac{V_0(y)}{2}-\log
y-\int_{\lambda_3}^y
\frac{(s^2-\lambda_1^2)\sqrt{(s^2-\lambda_2^2)(s^2-\lambda_3^2)}}{2s^3}ds\right).
\end{equation}
The function $g(y)$ is analytic in
$\mathbb{C}\setminus(-\infty,\lambda_3)$ and $g(y) \sim \log y$ as
$y\to \infty$. Now, let $\Sigma: = \Sigma_1 \cup \Sigma_2$, where
\begin{equation}
  \label{eq:interval}
  \Sigma_1 := \left [-\lambda_3,-\lambda_2\right] \quad \text{and} \quad 
    \Sigma_2 := \left[\lambda_2,\lambda_3\right]. 
\end{equation}
We will dedicate the rest of this section to show that $g(y)$ 
satisfies the conditions \eqref{eq:ineqg} with
\begin{equation}
  \label{eq:support_measure}
  \supp\left(\mu\right) = \Sigma = \Sigma_1 \cup \Sigma_2.
\end{equation}

\begin{lemma}
\label{le:jumps}
The function $g(y)$ defined in equation~\emph{\eqref{eq:gfun}}
satisfies the  jump discontinuities
\begin{subequations}
\label{eq:jumps}
\begin{alignat}{2}
\label{eq:jump_1}
&g_+(y)+g_-(y)-V_0(y)=l, & & y\in\Sigma,\\
\label{eq:jump_2}
&g_+(y)-g_-(y)=2\pi i, & & y\in(-\infty,-\lambda_3),\\
\label{jump_3}
&g_+(y)-g_-(y)=\pi i,& \quad& y\in(-\lambda_2,\lambda_2).
\end{alignat}
\end{subequations}
\end{lemma}
\begin{proof} Since $V_0(y)/2$ has no jump
  discontinuities in $\mathbb{C}$, we only concentrate on the integral
  in the right-hand side of (\ref{eq:gfun}). Let us write 
\begin{subequations}
\label{eq:gtilde}
\begin{align}
  \label{eq:nu}
  \nu(y)  & :=\frac{(y^2-\lambda_1^2)
\sqrt{(y^2-\lambda_2^2)(y^2-\lambda_3^2)}}{2y^3},\\
  \label{eq:gtilde_2}
   \tilde{g}(y) & :=\int_{\lambda_3}^y \nu(s)ds.
\end{align}
\end{subequations}
If $y \in \Sigma_2$, then
\begin{equation}\label{eq:bound1}
\begin{split}
\tilde{g}_+(y)+\tilde{g}_-(y)
=\int_{\lambda_3}^y\bigl(\nu_+(s)+\nu_-(s)\bigr)ds.
\end{split}
\end{equation}
Since $\nu(y)$ changes sign across $\Sigma$, we
have
\begin{equation}
\label{eq:jumpsq}
\nu_+(y)=-\nu_-(y),\quad y\in\Sigma.
\end{equation}
Equations~\eqref{eq:bound1} and~\eqref{eq:jumpsq}
imply that~\eqref{eq:jump_1} is satisfied for $y \in \Sigma_2$.

Suppose now that $y\in \Sigma_1$. Let
$\Gamma_{\pm}$ be a contour that consists of 3 parts: the first one
$\Gamma_{\pm}^1$ goes from $\lambda_3$ to $\lambda_2$ on the
positive/negative side of the real axis; the second part
$\Gamma_{\pm}^2$ is a semicircle from $\lambda_2$ to $-\lambda_2$
in the upper/lower half plane; the last part $\Gamma_{\pm}^3$ goes from
$-\lambda_2$ to $y$ on the positive/negative side of the real axis.
Then, the jump on $\Sigma_1$ is given by
\begin{equation}
\label{eq:bound2}
\begin{split}
\tilde{g}_+(y)+\tilde{g}_-(y)&=\int_{\Gamma_+}\nu(s)ds +
\int_{\Gamma_-}\nu(s)ds = \int_{-\lambda_2}^{y}\bigl(\nu_+(s)+\nu_-(s)\bigr)ds\\
& \quad + \int_{\lambda_3}^{\lambda_2}\bigl(\nu_+(s)+\nu_-(s)\bigr)ds
 +\int_{\Gamma_{+}^2}\nu(s)ds+\int_{\Gamma_-^2}\nu(s)ds.
\end{split}
\end{equation}
Since $\nu_+(s)=-\nu_-(s)$ on $\Sigma$, the first two terms in the
right-hand side of equation~\eqref{eq:bound2} are zero. Furthermore,
under the map $s\mapsto -s$, the contour $\Gamma_+^2$ becomes
$-\Gamma_-^2$ and $\nu(-s)=-\nu(s)$. Hence, the sum of the last two
terms are zero too. This proves \eqref{eq:jump_1} on $\Sigma_1$.

Let  $y \in (-\infty,-\lambda_3)$.  We simply have
\begin{equation}
  \label{eq:res_theo}
  \tilde g_+(y) - \tilde g_-(y) = 2\pi i \res_{y=\infty} \tilde g(y). 
\end{equation}
Then, equation~\eqref{eq:jump_2} follows
from 
\begin{equation}
  \label{eq:res_infity}
  \res_{y = \infty} \tilde g(y) = -\frac{\lambda_1^2 +
    \left(\lambda_2^2 + \lambda_3^2\right)/2}{2} =-1,
\end{equation}
where we have used~\eqref{eq:aj_1}.

Let $y \in (-\lambda_2,\lambda_2)$ and consider the contours
$\tilde \Gamma_{\pm}:= \Gamma^1_\pm \cup \tilde \Gamma^2_\pm$ and
$\tilde \Gamma^3_{\pm}$. Now $\tilde \Gamma^2_{\pm}$ joins $\lambda_2$
to $y$ and and $\tilde \Gamma^3_{\pm}$ connects $-\lambda_2$ to
$-\lambda_3$ respectively. As before $\pm$ denotes the upper/lower
half plane respectively.  We have
\begin{equation}
  \label{eq:jump_4}
  \tilde g_+(y) - \tilde g_-(y)  = \int_{\tilde \Gamma_+} \nu(s)ds -
   \int_{\tilde \Gamma_-} \nu(s)ds = \int_{\Gamma^1_+ \cup
     \left(-\Gamma_-^1\right)}\nu(s)ds,
  \end{equation}
where we have used
\begin{equation}
  \label{eq:res_at_zero}
  \res_{y=0}\tilde g(y) = \frac{\lambda_1^{-2} +
    \left(\lambda_2^{-2} + \lambda_3^{-2}\right)/2}{2}=0,
\end{equation}
which follows from equation~\eqref{eq:aj_2}.
The function $\nu(s)$ is odd and $\Gamma_{\pm}^1$ is mapped into
$-\tilde \Gamma^3_{\mp}$ by $ s \mapsto - s$, therefore
\begin{equation}
  \label{eq:symmetry}
  \int_{\tilde \Gamma^3_+ \cup \left(-\tilde
      \Gamma_-^3\right)}\nu(s)ds =
\int_{\Gamma^1_+ \cup \left(-\Gamma_-^1\right)}\nu(s)ds.
\end{equation}
Finally, Cauchy's theorem gives
\begin{equation}
  \label{eq:caucjy}
  \int_{\Gamma^1_+ \cup \left(-\Gamma_-^1\right)}\nu(s)ds =   
   \pi i \res_{y=\infty} \tilde{g}(y) =  -i\pi.
\end{equation}
This completes the proof of equation~\eqref{jump_3}.
\end{proof}

We are now ready to prove that the definition~\eqref{eq:gfun} gives
the \textit{g}-function. 
\begin{proposition}
\label{pro:measure}
Suppose that $\lambda_1$, $\lambda_2$ and $\lambda_3$ satisfy the
conditions of Lemma~\emph{\ref{le:realA}}, that
$\ipart\left(\lambda_1\right) >0$ and $0 < \lambda_2 <
\lambda_3$. Then, the function $g(y)$ defined in
\emph{\eqref{eq:gfun}} satisfies
\begin{subequations}
\label{eq:ineqg2}
\begin{alignat}{2}
\label{eq:ineqg2a}
&g_+(y)+g_-(y)-V_0(y)=l,&\qquad& y\in\Sigma,\\
\label{eq:ineqg2b}
&\mathrm{Re}\left(g_+(y)+g_-(y)\right)-V_0(y)< l, & &
y\in\mathbb{R}\setminus \Sigma.
\end{alignat}
\end{subequations}
\end{proposition}
\begin{proof}
  Equation~\eqref{eq:ineqg2a} was proven in Lemma~\ref{le:jumps}.  We
  are left to prove inequality~\eqref{eq:ineqg2b}. From the jump
  discontinuities~\eqref{eq:jumps}, we see that outside $\Sigma$ the real
  parts of $g_+(y)$ and $g_-(y)$ are equal. In particular, the real
  part of $g(y)$ is continuous outside $\Sigma$.

Equation~\eqref{eq:jump_1} and
\begin{equation}
\label{eq:g_gt_rel}
\tilde{g}(y)=\frac{V_0(y)}{2}-g(y)+\frac{l}{2}
\end{equation}
give the equality
\begin{equation}
\label{eq:gtildesign}
\rpart\bigl(\tilde{g}_+(y)+\tilde{g}_-(y)\bigr)=0,\quad
y\in\Sigma.
\end{equation}
Now note that 
\begin{equation}
\label{eq:nusign}
\begin{cases}
\nu(y)>0& \text{if $y\in(-\lambda_2,0)\cup(\lambda_3,\infty)$,} \\
\nu(y)<0& \text{if $y\in(-\infty,-\lambda_3)\cup(0,\lambda_2)$}.
\end{cases}
\end{equation}
Thus, $\rpart\tilde{g}(y)$ is an increasing function in
$(-\lambda_2,0)\cup(\lambda_3,\infty)$ and decreases in
\newline
$(-\infty,-\lambda_3)\cup(0,\lambda_2)$.  This property and
\eqref{eq:gtildesign} imply
\begin{equation}
\label{eq:last_ineq}
\begin{split}
\rpart\bigl(\tilde{g}_+(y)+\tilde{g}_-(y)\bigr)>0, \quad
x\in\mathbb{R}\setminus \Sigma.
\end{split}
\end{equation}
The proposition now follows from equations~\eqref{eq:g_gt_rel}
and~\eqref{eq:last_ineq}.
\end{proof}
\section{The Riemann-Hilbert analysis}
\label{se:RH}
The purpose of this section is to study the asymptotic limit of the
solution $Y(y)$ of the R-H problem~\eqref{eq:RHP} using the nonlinear
steepest descent analysis developed by Deift \textit{et
  al}~\cite{DKV,DKV2}.  One of the main ingredients is the
\textit{g}-function computed in \S\ref{se:equilibrium_measure}.

\subsection{Deformation of the Riemann-Hilbert problem}

Let us introduce 
\begin{equation}
\label{eq:Fx}
F(y):=\frac{v_1 q(y)}{2\pi
i}\left(\int_{\Sigma}\frac{ds}{sq_+(s)(s-y)}+
\xi\int_{-\lambda_2}^{\lambda_2}\frac{ds}{q(s)(s-y)}\right),
\end{equation}
where $q(y)$ and $\xi$ are defined by
\begin{subequations}
\label{eq:qxi}
\begin{align}
\label{eq:q}
q(y) & :=\sqrt{(y^2-\lambda_2^2)(y^2-\lambda_3^2)},\\
\label{eq:K0}
K_0& :=2\int_{\lambda_2}^{-\lambda_2}\frac{ds}{q(s)},\\
\label{eq:xi}
\xi & :=-\frac{\int_{\Sigma}\frac{ds}{sq_+(s)}}%
{\int_{-\lambda_2}^{\lambda_2}\frac{ds}{q(s)}}
=-\frac{2\pi i}{K_0\lambda_2\lambda_3}.
\end{align}
\end{subequations}
The right-hand side of equation~\eqref{eq:xi} follows from the residue theorem.

The function $F(y)$ is bounded at the points $\pm \lambda_2$, $\pm
\lambda_3$ and satisfies the following scalar R-H problem:
\begin{equation}\label{eq:RHPF}
\begin{alignedat}{2}
1. \quad & \text{$F(y)$ is analytic in 
$\mathbb{C}\setminus[-\lambda_3,\lambda_3]$,} &  &\\
2. \quad & F_+(y)+F_-(y)=\frac{v_1}{y},& & y\in\Sigma,\\
3. \quad & F_+(y)-F_-(y)=\xi v_1,& & y\in(-\lambda_2,\lambda_2),\\
4. \quad & F(y)=O(y^{-1}),&  & y\rightarrow\infty.
\end{alignedat}
\end{equation}
Let us define
\begin{equation}
\label{eq:Tx}
T(y): =e^{\left(-\frac{Nl}{2}\right)\sigma_3}Y(y)
e^{-\bigl(Ng(y)-F(y)\bigr)\sigma_3}e^{\frac{Nl\sigma_3}{2}},
\end{equation}
where $\sigma_3=\left(\begin{smallmatrix} 
1 &0  \\ 0 & -1 \end{smallmatrix}\right)$.
The matrix function $T(y)$ is the solution to the R-H problem
\begin{equation}
\begin{alignedat}{2}
1. \quad & \text{$T(y)$ is analytic in 
              $\mathbb{C}\setminus\mathbb{R}$} & & \\
2. \quad & T_+(y)=T_-(y)J_T(y), & & y\in\mathbb{R},\\
3. \quad & T(y)=I+O(y^{-1}), & \quad & y\rightarrow\infty,
\end{alignedat}
\end{equation}
where 
\begin{equation}
J_T(y) :=\begin{pmatrix} 
e^{-N\bigl(g_+(y)-g_-(y)\bigr) +F_+(y)-F_-(y)} & 
e^{-N\bigl(\tilde{g}_+(y)+\tilde{g}_-(y)\bigr) - F_{+}(y) -F_{-}(y) 
+\frac{v_1}{y}}  \\
0 &e^{N\bigl(g_+(y)-g_-(y)\bigr)  -F_+(y)+F_-(y)}
\end{pmatrix},\quad y\in\mathbb{R}
\end{equation}
and $\tilde{g}(y)$ is defined in \eqref{eq:gtilde_2}.

\begin{figure}[ht]
 \psfrag{C_1}[1.5][0.0]{\small$\Xi_{+ j}$}
 \psfrag{L_+}[1.5][0.0]{\small$L_{+ j}$}
 \psfrag{L_-}[1.5][0.0]{\small$L_{- j}$}
 \psfrag{C_2}[1.5][0.0]{\small$\Xi_{- j}$} 
\centering
\includegraphics[scale=0.3]{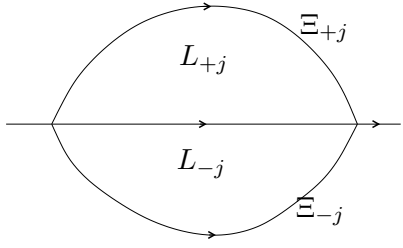}
\caption{The opening of the lenses around the intervals $\Sigma_j$,
  $j=1,2$. $L_{\pm  j}$ and $\Xi_{\pm j}$ are the interiors and
  contours of the lenses in the upper/lower half plane
  respectively. We also denote $\Xi_j=\Xi_{+ j}\cup \Xi_{- j}$ and
  $\Xi=\Xi_{1}\cup \Xi_{2}$, $j=1,2$.}
\label{fig:lens}
\end{figure}

We now perform a standard technique in the steepest decent analysis
(see \cite{BI,DKV,DKV2}): we open lenses around the intervals
$\Sigma_j$, $j=1,2$.  The interiors $L_{\pm j}$ and contours
$\Xi_{\pm j}$ of the lenses are defined as in figure \ref{fig:lens}.

Let us introduce the matrix function
\begin{equation}
\label{eq:S}
S(y): = \begin{cases}
         T(y), & y \in \mathbb{C} \setminus \left(L_{+ j}\cup L_{- j}\right), \\
         T(y)\begin{pmatrix} 1 & 0  \\
-e^{2N\tilde{g}(y) + 2F(y) - \frac{v_1}{y}} &1
\end{pmatrix}, & \text{$y\in L_{+ j}$,} \\
         T(y)\begin{pmatrix} 1 & 0  \\
e^{2N\tilde{g}(y) +  2F(y) - \frac{v_1}{y}} &1
\end{pmatrix}, & \text{$y\in L_{- j}$.}
       \end{cases}
\end{equation}
for $j=1,2$. It satisfies the R-H problem
\begin{equation}
\label{eq:RHS}
\begin{alignedat}{2}
1. \quad &\text{$S(y)$ is analytic in
  $\mathbb{C}\setminus\mathbb{R}$,} &&\\
2. \quad & S_+(y)=S_-(y)J_S(y),& \quad &  y\in\mathbb{R},\\
3. \quad & S(y)=I+O(y^{-1}),& \quad&  y\rightarrow\infty.
\end{alignedat}
\end{equation}
where $J_S(y)$ is the jump matrix
\begin{equation}
\label{eq:Sjump1}
J_S(y):=\begin{cases}
\begin{pmatrix} 1 & 0  \\
e^{2N\tilde{g}(y)+ 2F(y)-\frac{v_1}{y}} &1
\end{pmatrix}, & y \in \Xi,\\
\begin{pmatrix} 0 & 1  \\
-1 &0\end{pmatrix}, &  y\in \Sigma,\\
\begin{pmatrix} 1 & e^{-2N\tilde{g}(y) - 2F(y)+\frac{v_1}{y}}  \\
0 &1\end{pmatrix}, &  x\in \mathbb{R}\setminus(-\lambda_3,\lambda_3),\\
\begin{pmatrix} e^{N \pi i + \xi v_1} & 
e^{-N\bigl(\tilde{g}_+(y)+\tilde{g}_-(y)\bigr) -F_+(y)-F_-(y)+\frac{v_1}{y}}  \\
0 &e^{-N\pi i - \xi v_1}\end{pmatrix}, &y\in (-\lambda_2,\lambda_2)
\end{cases}
\end{equation}
The conditions (\ref{eq:ineqg2}) imply that away from some small
discs $D_{\pm \lambda_j}$ of radius $\delta$ centered at $\pm\lambda_j$,
$j=2,3$, the off-diagonal entries of the jump matrix $J_S(y)$
are exponentially small in $N$, except on the intervals $\Sigma_1$ and
$\Sigma_2$. This suggests the following approximation to the R-H
problem for $S(y)$:
\begin{equation}
\label{eq:sinf}
\begin{alignedat}{2}
  1. \quad & \text{$S^{\infty}(y)$ is analytic in 
  $\mathbb{C}\setminus[-\lambda_3,\lambda_3]$}, &&\\
  2. \quad & S_+^{\infty}(y)=S_-^{\infty}(y)J^{\infty}(y), && 
    y\in [-\lambda_3,\lambda_3],\\
  3. \quad & S^{\infty}(y)=I+O(y^{-1}),& \quad & y\rightarrow\infty,
\end{alignedat}
\end{equation}
where 
\begin{equation}
\label{eq:tildejinf}
J^{\infty}(y):= \begin{cases}
\begin{pmatrix} e^{N \pi i + \xi v_1} & 0  \\
0 &e^{- N\pi i - \xi v_1 }\end{pmatrix}, & y\in(-\lambda_2,\lambda_2),\\
\begin{pmatrix} 0 & 1  \\
-1 &0\end{pmatrix}, & y \in\Sigma.
\end{cases}
\end{equation}
The approximation $S^\infty(y)$ is known as \textit{outer parametrix.}

\subsection{The outer parametrix}
\label{se:outer}
Here and in the rest of
\S\ref{se:RH}, $\pm \lambda_j$ will always refer to the edge points of
the support $\Sigma$ of the equilibrium measure. Hence, $j=2,3$ only.

The solution to the R-H problem~\eqref{eq:sinf} exists and is
uniformly bounded in $N$ outside of small discs $D_{\pm \lambda_j}$
around the points $\pm\lambda_j$.   Such a solution can be
constructed in terms of  elliptic theta functions as in Deift
\textit{et al}~\cite{DKV2}.  Here we follow their treatment.

\begin{figure}[ht]
\centering 
 \psfrag{a}[1.25][0.0]{\small$a$}
 \psfrag{b}[1.25][0.0]{\small$b$}
 \psfrag{x+sigma_t}[1.25][0.0]{\small$\lambda_3$}
 \psfrag{alpha_t}[1.25][0.0]{\small$-\lambda_3$}
 \psfrag{beta_t}[1.25][0.0]{\small$-\lambda_2$}
 \psfrag{x-sigma_t}[1.25][0.0]{\small$\lambda_2$}
\includegraphics[scale=0.75]{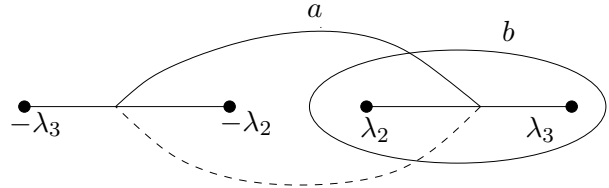}
\caption{The $a$ and $b$ cycle of the elliptic curve
(\ref{eq:Lie}).}\label{fig:cycle}
\end{figure}

Let $\Lie$ be the elliptic curve 
\begin{equation}\label{eq:Lie}
q^2=(y^2-\lambda_2^2)(y^2-\lambda_3^2)
\end{equation}
and choose a canonical basis of cycles as in figure
\ref{fig:cycle}.
Then, the holomorphic 1-form $\omega(y)$ dual to this set of cycles is
given by
\begin{equation}\label{eq:1form}
\omega(y):=\frac{dy}{K_0q(y)},
\end{equation}
where $K_0$ was introduced in equation~\eqref{eq:K0}. The Abel map is
defined by
\begin{equation}\label{eq:Abel}
u(y):=\int_{\lambda_3}^{y}\frac{ds}{K_0q(s)}.
\end{equation}
where the contour of integration is chosen such that it does not
intersect the interval $(-\infty,\lambda_3)$. Let $\Pi$ be the
$b$-period of the 1-form $\omega(y)$:
\begin{equation}\label{eq:bperiod}
\Pi:=2\int_{\lambda_3}^{\lambda_2}\frac{dy}{K_0q_+(y)}.
\end{equation}
The elliptic theta function for the curve (\ref{eq:Lie}) with this
choice of cycles is given by
\begin{equation}
\label{eq:thetadef}
\theta(s): =\sum_{m\in\mathbb{Z}}e^{i\pi \Pi m^2 +2\pi i sm},
\end{equation}
Consider the function
\begin{equation}\label{eq:gamma}
\gamma: =\left(\frac{(y-\lambda_2)(y+\lambda_3)}{(y+\lambda_2)(y-\lambda_3)}\right)^{\frac{1}{4}},
\end{equation}
where the arguments of the individual factors in the fourth root are
chosen to be between $-\pi$ and $\pi$. Then the solution
to~(\ref{eq:sinf}) is given by
\begin{subequations}
\label{eq:para}
\begin{align}
\label{eq:para_1}
S^{\infty}(y)&:=H\begin{pmatrix}\frac{\gamma+\gamma^{-1}}{2}
\frac{\theta\left(u(y)-\frac{N}{2} -\frac{v_1\xi}{2\pi
i}+d\right)}{\theta\left(u(y)+d\right)}
&\frac{\gamma-\gamma^{-1}}{-2i}\frac{\theta
\left(-u(y)-\frac{N}{2} - \frac{v_1\xi}{2\pi
i}+d\right)}{\theta\left(-u(y)+d\right)}\\
\frac{\gamma-\gamma^{-1}}{2i}\frac{\theta\left(u(y)-\frac{N}{2} - 
\frac{v_1\xi}{2\pi
i}-d\right)}{\theta(u(y)-d)}&
\frac{\gamma+\gamma^{-1}}{2}\frac{\theta(-u(y)-\frac{N}{2}-\frac{v_1\xi}{2\pi
i}-d)}{\theta(u(y)+d)}\end{pmatrix},\\
\label{eq:para_2}
H&:=\diag\left(\frac{\theta(u(\infty)+d)}{\theta\left(
u(\infty)-\frac{N}{2}-\frac{v_1\xi}{2\pi
i}+d\right)},\frac{\theta\left(u(\infty)+d\right)}%
{\theta\left(-u(\infty)-\frac{N}{2}-\frac{v_1\xi}{2\pi
i}+d\right)}\right).
\end{align}
\end{subequations}
where $d$ is the constant 
\begin{equation}
d:=-\frac{1}{2}-\frac{\Pi}{2}+u_+(0)=-\frac{1}{2}-\frac{\Pi}{2}+
\int_{\lambda_3}^{\lambda_2}\frac{ds}{K_0q_+(s)}
+\int_{\lambda_2}^0\frac{ds}{K_0q_+(s)}.
\end{equation}
Using the definition of $b$-period \eqref{eq:bperiod}, of
$K_0$~\eqref{eq:K0} and the fact that $q(-s)=q(s)$ give
$d=-\frac{1}{4}$.

\subsection{Local parametrices near $\pm \lambda_2$ and $\pm
  \lambda_3$ }
\label{se:local} 

Near the edge points $\pm \lambda_j$ the approximation of
$S(y)$ by $S^{\infty}(y)$ fails.  Therefore, we must solve the R-H
problem~\eqref{eq:RHS} in small neighborhoods of these points and
match the solutions to the outer parametrix~(\ref{eq:sinf}) up to an
error term of order $O(N^{-1})$. More precisely, let $\delta>0$ and 
$D_{\pm \lambda_j}$ be a disc of radius $\delta$ centered at 
$\pm \lambda_{ j}$. We would like to construct local
parametrices $S^{\left(\pm \lambda_j\right)}(y)$ in $D_{\pm \lambda_j}$ such that
\begin{equation}
\label{eq:localpara}
\begin{alignedat}{2}
  1.\quad & \text{$S^{\left(\pm \lambda_j\right)}(y)$ is analytic in $D_{\pm \lambda_j}\setminus
    \bigl(D_{\pm \lambda j}\cap\left(\mathbb{R}\cup\Xi\right)\bigr)$,} &&\\
  2.\quad &S_+^{\left(\pm \lambda_j\right)}(y)=S_-^{\left(\pm \lambda_j\right)}(z)J_S(y), && 
   y\in D_{\pm \lambda_j}\cap\left(\mathbb{R}\cup\Xi\right),\\
  3.\quad &S^{\left(\pm \lambda_j\right)}(y)=\left(I+O(N^{-1})\right)S^{\infty}(y), & &  y\in\p D_{\pm \lambda_j}.
\end{alignedat}
\end{equation}
These local parametrices are given by
\begin{equation}
  \label{eq:loc_para_ex}
S^{\left(\pm \lambda_j\right)}(y):=E_n^{\left(\pm \lambda_j\right)}(y)\Psi^{\left(\pm \lambda_j\right)}(\zeta_{\pm
j})e^{\bigl(N\tilde{g}(y) + F(y)-\frac{v_1}{2y}\bigr)\sigma_3}.
\end{equation}
The matrix functions $\Psi^{\left(\pm \lambda_j\right)}$ are
constructed using Airy functions (see, \textit{e.g.},~\cite{D}
pp. 213--216).  The explicit expression of $\Psi^{\left(\pm
    \lambda_j\right)}$ is quite lengthy.  Besides, it does not enter
in our calculations. Therefore, we refer the interested reader to the
original literature (see~\cite{BI,D,DKV,DKV2}).  The $\zeta_{\pm
  \lambda_j}$'s are conformal maps inside the neighborhoods $D_{\pm
  \lambda_j}$ given by the analytic continuation of the functions
$\zeta_{\pm \lambda_j}^+$ to the whole $D_{\pm \lambda_j}$:
\begin{equation}
\label{eq:zeta}
\zeta_{\pm \lambda_j}^{+}:=\left(\frac{3}{2}N\right)^{\frac{2}{3}}
\bigl(\tilde{g}(y)-\tilde{g}(\lambda_{\pm
j})\bigr)^{\frac{2}{3}},\quad \mathrm{Im}(y)>0.
\end{equation}
The matrix $E_n^{\left(\pm \lambda_j\right)}(y)$ is invertible and is the analytic
continuation to the whole  $D_{\pm \lambda_j}$ of the following quantity:
\begin{equation}
E_{n,+}^{(\pm \lambda_j)}(y)
:=\sqrt{\pi}e^{-\frac{i\pi}{12}}S^{\infty}_+(y)
e^{\left(-F(y)+ \frac{v_1}{2y}\right)
\sigma_3}e^{\frac{i\pi}{4}\sigma_3}\begin{pmatrix}1&\mp
(-1)^{j+1}\\\pm (-1)^{j+1}&1\end{pmatrix}\left(\pm\zeta_{\pm
\lambda_j}\right)^{\frac{\sigma_3}{4}}.
\end{equation}

\begin{remark} 
Since $E_n^{\left(\pm \lambda_j\right)}(y)$ is analytic inside $D_{\pm
  \lambda_j}$, we see that near the points $\pm\lambda_j$, the
function $S^{\infty}(y)$ behaves like
\begin{equation}
\label{eq:Ssing}
\begin{split}
S^{\infty}(y)& \sim S_0^{\left(\pm \lambda_j\right)}(y)(y\mp\lambda_j)^{-\frac{\sigma_3}{4}}
\begin{pmatrix}1&\pm (-1)^{j+1}\\
\mp
(-1)^{j+1}&1\end{pmatrix}\\
& \quad \times e^{-\frac{i\pi}{4}\sigma_3}
e^{\left(-F(y)+\frac{v_1}{2y}\right)\sigma_3},\quad
y\to\pm\lambda_j, 
\end{split}
\end{equation}
where $S_0^{\left(\pm \lambda_j\right)}(y)$ is holomorphic and invertible at
$\pm\lambda_j$.
\end{remark}
\subsection{The final transformation of the Riemann-Hilbert
  problem}\label{se:final} 
We now show that the parametrices we
constructed in \S\ref{se:outer} and \S\ref{se:local} are
indeed good approximations to the solution $S(y)$ of the
R-H problem~(\ref{eq:RHS}).

\begin{figure}[ht]
\centering 
\psfrag{l2}[1.25][0.0]{\small$\lambda_2$}
\psfrag{l3}[1.25][0.0]{\small$\lambda_3$}
\psfrag{-l2}[1.25][0.0]{\small$-\lambda_2$}
\psfrag{-l3}[1.25][0.0]{\small$-\lambda_3$}
\includegraphics[scale=0.75]{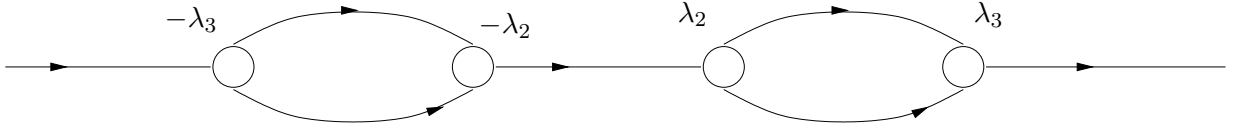}
\caption{The contour $\Gamma_R$.}\label{fig:sigma}
\end{figure}

Let us define 
\begin{equation}
\label{eq:Rx}
R(y):= \begin{cases}
         S(y)\left(S^{\left(\pm \lambda_j\right)}(y)\right)^{-1}, & y 
        \in D_{\pm \lambda_j}, \\
         S(y)\left(S^{\infty}(y)\right)^{-1}, & y \in D_{\pm \lambda_j}.
       \end{cases}
\end{equation}
Then the function $R(y)$ has jump discontinuities on the contour
$\Gamma_R$ shown in figure \ref{fig:sigma}.
In particular, $R(y)$ satisfies the R-H problem
\begin{equation}\label{eq:RHR}
\begin{alignedat}{2}
1. \quad & \text{$R(y)$ is analytic in $\mathbb{C}\setminus\Gamma_R$,} &&\\
2.\quad & R_+(y)=R_-(y)J_R(y), && y \in \Gamma_R,\\
3. \quad & R(y)=I+O(y^{-1}), & \quad & y \to \infty.
\end{alignedat}
\end{equation}
From the definition of $R(y)$ it follows that 
the jump matrix $J_R(y)$ has the following order of magnitude as $N
\to \infty$:
\begin{equation}
\label{eq:Jx}
J_R(y)=
       \begin{cases}
         I+O(N^{-1}), & y\in\p D_{\pm \lambda_j}, \\
         I+O\left(e^{-N\eta}\right), & \text{for some fixed $\eta>0$
         for  $y \in  \Gamma_R \setminus D_{\pm \lambda_j}$.}
       \end{cases}
\end{equation}
Then, using well established techniques (see,\textit{e.g}~\cite{DKV}
\S7) we obtain
\begin{equation}
\label{eq:Rest}
\begin{split}
R(y)=I+O\left(\frac{1}{N\left(\abs{y}+1\right)}\right),
\end{split}
\end{equation}
uniformly in $\mathbb{C}$. Therefore, the solution $S(y)$ of the R-H
problem (\ref{eq:RHS}) can be approximated by $S^{\infty}(y)$ and
$S^{\left(\pm \lambda_j\right)}(y)$:
\begin{equation}
\label{eq:approxS}
S(y)= \begin{cases}
  \left(I+O\left(N^{-1}\right)\right)S^{\left(\pm \lambda_j\right)}(y), & y \in
  D_{\pm \lambda_j}, \\
  \left(I+O\left(N^{-1}\right)\right)S^{\infty}(y), & y \in 
  \mathbb{C} \setminus D_{\pm \lambda_j}.
       \end{cases}
\end{equation}
Combining these expressions with equations~\eqref{eq:S}
and~\eqref{eq:Tx} we obtain an asymptotic formula for the solution of
the original R-H problem~\eqref{eq:RHP}, which can be inserted in the
differential identities~\eqref{eq:diffid}.

\section{Asymptotics of   the differential identities~\eqref{eq:diffid}}
\label{se:asy_diffid}
In this section we will compute an asymptotic formula for the
partition function $G_N(v_1,v_2)$ defined in \eqref{eq:tildeB}. The
analysis is similar to those ones carried out in
\cite{IJK1,IJK2,IMM}. 
\subsection{The finite $v_2$ regime} 
Consider the trace in the
integrals~\eqref{eq:diffid}: 
\begin{equation}
\label{eq:alpha}
\alpha(y): =\tr\left(Y^{-1}(y)Y^{\prime}(y)\sigma_3\right).
\end{equation}
Since $\alpha(y)$ is analytic in a neighbourhood of the origin, in
order to compute the integrals~\eqref{eq:diffid} we only need the
first two terms in its Taylor expansion at $y=0$.  We have
\begin{equation}
\alpha(y) = \tr\left(S^{-1}(y)S^{\prime}(y)\sigma_3\right)
+ 2Ng^{\prime}(y) - F^{\prime}(y).
\end{equation}
By using $S(y)=R(y)S^{\infty}(y)$ and the expression (\ref{eq:para})
for $S^{\infty}(y)$ we obtain
\begin{equation}
\label{eq:alpha1}
\alpha(y)= \tr\left(\Theta^{-1}(y) \Theta^{\prime}(y)\sigma_3\right)
+ 2N g^{\prime}(y)-F^{\prime}(y)+ O\left(\frac{1}{N}\right),
\end{equation}
where 
\begin{equation}
  \label{eq:Theta_def}
   \Theta(y)=H^{-1}S^{\infty}(y).
\end{equation}
It is convenient to split the computation of the right-hand side of
this equation in two parts.  We first determine
\begin{equation}
  \label{eq:trace_Theta}
  \alpha_0(y):=\tr\left(\Theta^{-1}(y) \Theta^{\prime}(y)\sigma_3\right),
\end{equation}
then we evaluate
\begin{equation}
  \label{eq:gp_fp}
  2Ng^{\prime}(y) - F^{\prime}(y).
\end{equation}

\begin{lemma}
\label{lem:alpha}
We have
\begin{equation}\label{eq:alpharoot}
\alpha_0(y)=\frac{M}{q(y)},
\end{equation}
where $q(y)$ is defined in \emph{\eqref{eq:Lie}} and $M$ is a constant.
\end{lemma}
\begin{proof} 
We first show that $\alpha_0(y)$ is analytic in $\mathbb{C}\setminus
\Sigma$ and that
\begin{equation}
  \label{eq:alpha_jump1}
   \alpha_{0,+}(y) = - \alpha_{0,-}(y), \quad  y \in \Sigma.
\end{equation}

By differentiating the jump conditions in \eqref{eq:tildejinf}, we see
that $\Theta^{\prime}(y)$ has the same jumps as $\Theta(y)$.
Therefore, the quantity $\Theta^{-1}(y)\Theta^{\prime}(y)$ has the
discontinuity 
\begin{equation}
\label{eq:jumptheta}
\left(\Theta^{-1}(y)\Theta^{\prime}(y)\right)_+=\bigl(J^{\infty}(y)\bigr)^{-1}
\bigl(\Theta^{-1}(y)\Theta^{\prime}(y)\bigr)_-J^{\infty}(y),
\quad y \in (-\lambda_3,\lambda_3)
\end{equation}
and is analytic elsewhere. The
jump discontinuities of $J^{\infty}(y)$ in~(\ref{eq:tildejinf}) imply
equation~\eqref{eq:alpha_jump1} and that
\begin{equation}
  \label{eq:alpha_jump2}
  \alpha_{0,+}(x)=\alpha_{0,-}(y),\quad y \in (-\lambda_2,\lambda_2)
\end{equation}
Therefore, $\alpha_0(y)$ must be equal to $q(y)$ multiplied by a
rational function $r(y)$. Since $\alpha_0(x)$ is of order
$O\left(y^{-2}\right)$ as $y \to \infty$ and is analytic away from the
points $\pm \lambda_2$ and $\pm \lambda_3$, $r(y)$ can only have poles
at the edge points $\pm\lambda_j$, $j=2,3$.  From the behaviour of
$S^{\infty}(y)$ near $\pm\lambda_j$ (see
equation~\eqref{eq:Ssing}), we see that $\alpha_0(y)$ can at worst
have a singularity of order 1 at $\pm\lambda_j$. This implies that
$r(y)$ is analytic at the points $\pm\lambda_j$. Hence, it must be a
constant $M$. This proves the lemma.
\end{proof}

The constant $M$ in (\ref{eq:alpharoot}) can be found from the
asymptotic expansion of $\Theta(y)$ as $y \to \infty$.
\begin{lemma}
\label{eq:constant}
The function $\alpha_0(y)$ in \emph{\eqref{eq:alpharoot}} 
is given by
\begin{equation}
\label{eq:T}
\alpha_0(y)=-\frac{2\pi i }{q(y)K_0\xi} \frac{\p}{\p
v_1}\log\Bigl(\theta\bigl(u(\infty)+\varsigma -\tfrac{v_1\xi}{2\pi
i}\bigr)\theta\bigl(u(\infty)-\varsigma+\tfrac{v_1\xi}{2\pi
i}\bigr)\Bigr).
\end{equation}
where $\varsigma=-\frac{N}{2}-\frac{1}{4}$.
\end{lemma}
\begin{proof} From equations~\eqref{eq:sinf} and~\eqref{eq:Theta_def}
  the asymptotic expansion $\Theta(y)$ at $y=\infty$ has the form
 \begin{equation}
\label{eq:thetaasym}
\Theta(y)=H^{-1}+\frac{\Theta_1}{y}+O\left(y^{-2}\right),\quad
y\rightarrow\infty,
\end{equation}
where $H$ is the constant in (\ref{eq:para}). Hence,
\begin{equation}
\label{eq:a0}
\alpha_0(y)=-\tr\left(\frac{H\Theta_1}{y^2}\sigma_3\right)
+O\left(y^{-3}\right),\quad y\rightarrow\infty.
\end{equation}
The diagonal entries of $\Theta_1$ are 
\begin{subequations}
\label{eq:H_diag_entries}
\begin{align}
\left(\Theta_1\right)_{11}&=-\frac{H_{11}^{-1}}{K_0}
\left(\frac{\theta^{\prime}\bigl(u(\infty)-\frac{N}{2}-\frac{v_1\xi}{2\pi
i}-\frac{1}{4}\bigr)}{\theta\bigl(u(\infty)-\frac{N}{2}-\frac{v_1\xi}{2\pi
i}-\frac{1}{4}\bigr)}
-\frac{\theta^{\prime}\bigl(u(\infty)
-\frac{1}{4}\bigr)}{\theta\bigl(u(\infty)-\frac{1}{4}\bigr)}\right)
+\gamma_1H_{11}^{-1},\\
\left(\Theta_1\right)_{22}&=-\frac{H_{22}^{-1}}{K_0}
\left(\frac{\theta^{\prime}\bigl(u(\infty)+\frac{N}{2}+\frac{v_1\xi}{2\pi
i}-\frac{1}{4}\bigr)}{\theta\bigl(u(\infty)+\frac{N}{2}+\frac{v_1\xi}{2\pi
i}-\frac{1}{4})}-\frac{\theta^{\prime}\bigl(u(\infty)-\frac{1}{4}\bigr)}%
{\theta\bigl(u(\infty)-\frac{1}{4}\bigr)}\right)
+\gamma_1H_{22}^{-1},
\end{align}
\end{subequations}
where $\gamma_1$ is the coefficient of $y^{-1}$ in the expansion of
$\left(\gamma+\gamma^{-1}\right)/2$. Finally, equation~\eqref{eq:T}
follows by substituting~\eqref{eq:H_diag_entries} into \eqref{eq:a0}.
\end{proof}

We can now compute the logarithmic derivatives \eqref{eq:diffid}.
\begin{proposition}
  \label{pr:finte_v1v2}
  Let $v_2$ be of order $O(1)$ as $N \to \infty$ and define
\begin{equation}
\label{eq:C}
C :=\frac{2v_1\lambda_2\lambda_3}{K_0}
\int_{\lambda_2}^{-\lambda_2}\frac{ds}{s^2q(s)}.
\end{equation}
 Then,
  \begin{subequations}
 \label{eq:diffidasym}
\begin{align}
 \frac{\p \log G_N}{\p v_1}&=
 \frac{\pi i\lambda_1^2}{v_2K_0\xi}\frac{\p}{\p v_1}
\log\Bigl(\theta\bigl(u(\infty)+\varsigma-\tfrac{v_1\xi}{2\pi
i}\bigr)\theta\bigl(u(\infty)+\varsigma+\tfrac{v_1\xi}{2\pi
i}\bigr)\Bigr)  \nonumber  \\
&  \quad -\frac{v_1}{2\lambda_1^2}-\frac{C}{2\lambda_2\lambda_3}
+O\left(N^{-1}\right), \quad N \to \infty, \\
\frac{1}{N}\frac{\p \log G_N}{\p
v_2}& =N \left(\frac{1}{4}-\frac{v_2}{32}\left(\left(\frac{1}{\lambda_2^2}
-\frac{1}{\lambda_3^2}\right)^2
+\frac{8}{\lambda_1^4}\right)\right)+O\left(N^{-1}\right), \quad N \to
\infty.
\end{align}
\end{subequations}
\end{proposition}
\begin{proof}
  We are only left to determine the term~\eqref{eq:gp_fp}. First note
  that the derivative of $F(y)$ satisfies
\begin{equation}
\label{eq:RHPFprime}
\begin{alignedat}{2}
1. \quad & \text{$F^{\prime}(y)$ is analytic in $\mathbb{C}\setminus \Sigma$,}\\
2. \quad & F^\prime_+(y)+F^\prime_-(y)=-\frac{v_1}{y^2},& \quad &y\in\Sigma,\\
3. \quad & F^\prime(y)=O(y^{-2}),& & y\to \infty.
\end{alignedat}
\end{equation}
Moreover, since $F(y)$ is bounded at $\pm\lambda_j$, $j=2,3$,
$F^{\prime}(y)$ cannot have a singularity of order higher than 1 at
these points. A function with these properties must be of the form
\begin{equation}
\label{eq:Fprime}
F^{\prime}(y)=-\frac{v_1}{2y^2}-\frac{v_1\lambda_2\lambda_3-\tilde{C}y^2}%
{2y^2\sqrt{(y^2-\lambda_2^2)(y^2-\lambda_3^2)}},
\end{equation}
for some constant $\tilde{C}$. This means that
\begin{equation}
  \Omega(y):=\frac{v_1\lambda_2\lambda_3-\tilde{C}y^2}%
{2y^2\sqrt{(y^2-\lambda_2^2)(y^2-\lambda_3^2)}}dy
\end{equation}
is a meromorphic 1-form with a singularity at the origin such that
$\Omega(y) \sim v_1dy/2y^2$ as $y \to 0$ and is holomorphic
elsewhere. To determine the constant $\tilde{C}$, note that the jump
conditions (\ref{eq:RHPF}) implies that the $a$-period of $\Omega(y)$
must vanish. This gives $\tilde{C}=C$, where $C$ is defined in
equation~\eqref{eq:C}.

Therefore near $y=0$ the function $\alpha(y)$ behaves as
\begin{equation}
\begin{split}
  \alpha(y)&= N\Biggl(1-\frac{v_2}{8}
     \left(\left(\frac{1}{\lambda_2^2}
      -\frac{1}{\lambda_3^2}\right)^2
    + \frac{8}{\lambda_1^4}\right)\Biggr)y + \frac{v_1}{\lambda_1^2}
    + \frac{C}{\lambda_2\lambda_3}\\
     & \quad - \frac{2\pi i\lambda_1^2 }{v_2K_0\xi} \frac{\p}{\p
    v_1}\log\Bigl(\theta\left(u(\infty)+\varsigma-\tfrac{v_1\xi}{2\pi
      i}\right)\theta\left(u(\infty)+\varsigma+ \tfrac{v_1\xi}{2\pi
      i}\right)\Bigr)\\
  & \quad +O(y^2)+O(N^{-1}).
\end{split}
\end{equation}
Finally, inserting the right-hand side into
equations~\eqref{eq:diffid} gives formulae~\eqref{eq:diffidasym}.
\end{proof}

\subsection{The small $v_2$ regime}
When $v_2$ becomes  small, the local parametrices constructed in 
\S\ref{se:local} must be defined in neighborhoods $D_{\pm \lambda_2}$
whose radii $\delta$ is smaller than $\lambda_2$. This affects the
magnitude of the error term in the asymptotic formulae
(\ref{eq:diffidasym}). We now study its effect. We shall see
that by constructing local parametrices in shrinking neighborhoods
of the points $\pm\lambda_2$, we can extend the validity of
formulae (\ref{eq:diffidasym}) for $v_2 >N^{-3+\epsilon}$.

Since the error term is of order
\begin{equation}
\label{eq:error_term_tc}
O\left(\frac{\abs{S^{\infty}(y)}^2
e^{2\abs{F(y)}+\frac{v_1}{\abs{y}}}}%
{\abs{\zeta_{\pm 2}}^{\frac{3}{2}}}\right)
\end{equation}
on the boundary of $D_{\pm \lambda_2}$, we need to know the order of magnitude
of $S^{\infty}(y)$ and of the conformal map $\zeta_{\pm \lambda_2}$.  

\begin{proposition}
\label{pro:order}
The following asymptotic formulae as $v_2 \to 0$ hold.
\begin{enumerate}
\item The orders of magnitude of points $\lambda_1$ , $\lambda_2$ and
  $\lambda_3$ are 
\begin{subequations}
\label{eq:as_for_v2}
\begin{equation}
\label{eq:order_lambda}
\lambda_1=O\left(v_2^{\frac{1}{3}}\right),\quad
\lambda_2=O\left(v_2^{\frac{1}{3}}\right)\quad \text{and} \quad
\lambda_3=O\left(1\right).
\end{equation}
\item Let $D_{\pm \lambda_2}$ be small discs centered at $\pm\lambda_2$
with radii $\delta<\lambda_2$. Then, on the boundary of $D_{\pm \lambda_2}$ the
conformal map $\zeta_{\pm \lambda_2}$ defined in \emph{\eqref{eq:zeta}} is of
order
\begin{equation}
\label{eq:order_zeta}
\zeta_{\pm \lambda_2}=O\left(N^{\frac{2}{3}}v_2^{-\frac{1}{9}}\delta\right).
\end{equation}
\item There exists a constant $k=O(1)$ as $v_2 \to 0$ such that the
  outer parametrix $S^{\infty}(y)$ defined in
  equations~\emph{\eqref{eq:para}} is of order
\begin{equation}
\label{eq:order_out_param}
S^{\infty}(y)=O\left(\left \lvert \frac{\lambda_2}{\delta}\right 
\rvert^{\frac{1}{4}}e^{k\frac{v_1}{\lambda_2}}\right).
\end{equation}
\item The order of magnitude of the exponential in
  equation~\emph{\eqref{eq:error_term_tc}} is
\begin{equation}
  \label{eq:order_exp}
  \exp\left(2\abs{F(y)} + \tfrac{v_1}{\abs{y}}\right) 
= O\left(e^{k\frac{v_1}{\lambda_2}}\right).
\end{equation}
\end{subequations}
\end{enumerate}
\end{proposition}
\begin{proof}
  Equation~\eqref{eq:order_lambda} is an immediate consequence of
  formulae~(\ref{eq:A1}), (\ref{eq:A23-}) and (\ref{eq:sol23}). As $v_2
  \to 0$ we obtain
\begin{subequations}
\label{eq:orderlambda2}
\begin{align}
\lambda_1& =
\left(-2\right)^{-\frac{1}{6}}v_2^{\frac{1}{3}}+O\left(v_2\right),\\
\lambda_2& =\left(-\frac{1}{2}\right)^{\frac{1}{2}}
  \lambda_1+O\left(v_2\right),\\
\lambda_3& =2+O\left(v_2^{\frac{2}{3}}\right).
\end{align}
\end{subequations}
Next, consider the conformal maps $\zeta_{\pm \lambda_2}$. Inside
$D_{\pm \lambda_2}$ they behave as follows:
\begin{equation}
\zeta_{\pm \lambda_2}
=N^{\frac{2}{3}}\Bigl(\varphi_{\pm}(y-\lambda_2)+O\left((y-\lambda_2)^2
\right)\Bigr),
\end{equation}
where 
\begin{equation}
\label{eq:varphi}
\varphi_{\pm}:=\pm\frac{(\lambda_2^2-\lambda_1^2)^{\frac{2}{3}}
\Bigl(\lambda_2\left(\lambda_2^2-\lambda_3^2\right)\Bigr)^{\frac{1}{3}}}
{2^{\frac{1}{3}}\lambda_2^2}.
\end{equation}
Hence, $\zeta_{\pm \lambda_2}$ is of order
\begin{equation}
\zeta_{\pm \lambda_2}=O\left(N^{\frac{2}{3}}v_2^{-\frac{1}{9}}\delta\right)
\end{equation}
on the boundary of $D_{\pm \lambda_2}$.

Proving~\eqref{eq:order_out_param} requires more work. Firstly,
consider $\gamma\pm\gamma^{-1}$ in a small neighbourhood of
$\lambda_2$ . If $\abs{y-\lambda_2}=\delta$ is smaller than
$\lambda_2$, we have
\begin{equation}\label{eq:gammaasym}
\begin{split}
\gamma\pm\gamma^{-1}&=\left(\left(\frac{\delta}{2\lambda_2}\right)^{\frac{1}{4}}\left(\frac{\lambda_2+\lambda_3}{\lambda_2-\lambda_3}\right)^{\frac{1}{4}}
\pm\left(\frac{2\lambda_2}{\delta}\right)^{\frac{1}{4}}\left(\frac{\lambda_2-\lambda_3}{\lambda_2+\lambda_3}\right)^{\frac{1}{4}}\right)
\left(1+O\left(\frac{\delta}{\abs{\lambda_2}}\right)\right)\\
&=\pm\left(\frac{2\lambda_2}{\delta}\right)^{\frac{1}{4}}\left(1+O\left(\frac{\delta}{\abs{\lambda_2}}\right)+O\left(\lambda_2\right)\right).
\end{split}
\end{equation}
The case when $y=-\lambda_2$ can be treated analogously.

Let us now consider the theta functions that enter in the
definition~\eqref{eq:para}. First note that in the limit as
$\lambda_2\rightarrow 0$, the holomorphic 1-form $\omega(y)$ in
(\ref{eq:1form}) becomes a meromorphic 1-form with a simple pole at
$y=0$ with residue $1/(2\pi i)$ (see \cite{BBEIM} and 
\cite{IMM}). More explicitly
\begin{equation}
\label{eq:omegalim}
\omega(y)\to\frac{\lambda_3dy}{2\pi
y\sqrt{y^2-\lambda_3^2}}, \quad \lambda_2  \to 0.
\end{equation}
Furthermore, the constant $K_0$ has the limit
\begin{equation}
\label{eq:K0order}
K_0=\frac{2\pi}{\lambda_3}+O\left(\lambda_2^2\right), \quad \lambda_2
\to 0.
\end{equation}
By writing the period $\Pi$ as
\begin{equation}
\begin{split}
\Pi&=2\left(\int_{\lambda_3}^{\lambda_2^{\frac{1}{2}}}\frac{ds}{K_0q_+(s)}+
\int_{\lambda_2^{\frac{1}{2}}}^{\lambda_2}\frac{ds}{K_0q_+(s)}\right)\\
&=2\left(\int_{\lambda_3}^{\lambda_2^{\frac{1}{2}}}\frac{ds}{K_0s\sqrt{s^2-\lambda_3^2}}\bigl(1+O\left(\lambda_2\right)\bigr)+
\int_{\lambda_2^{\frac{1}{2}}}^{\lambda_2}\frac{ds}{K_0i\lambda_3\sqrt{s^2-\lambda_2^2}}\left(1+O\left(\lambda_2\right)\right)\right),
\end{split}
\end{equation}
we can compute its limit:
\begin{equation}
\label{eq:limpi}
\Pi=\left(\frac{1}{\pi i}\log|\lambda_2|-\frac{1}{\pi
i}\log|16\lambda_3^2|\right)\bigl(1+O\left(\lambda_2\right)\bigr),
\quad \lambda_2 \to 0.
\end{equation}
The Abel map $u(y)$ becomes
\begin{equation}
\label{eq:limabel}
u(y)=\int_{\lambda_3}^{\lambda_2}\omega(s)+\int_{\lambda_2}^y\omega(s)=\frac{\Pi}{2}
+\int_{\lambda_2}^y\omega(s).
\end{equation}
Using this expression we obtain
\begin{equation}
\label{eq:limitAbel}
\begin{split}
u(y)&=\frac{\Pi}{2}+
\left(\frac{\delta}{2\lambda_2}\right)^{\frac{1}{2}}\frac{2}{K_0
\sqrt{\lambda_2^2-\lambda_3^2}}\left(1+O\left(\frac{\delta}{\lambda_2}\right)
+O(\lambda_2)\right)\\
&=\frac{\Pi}{2}+\frac{1}{i\pi}\left(\frac{\delta}{2
\lambda_2}\right)^{\frac{1}{2}}\left(1+O\left(\frac{\delta}{\lambda_2}\right)+O(\lambda_2)\right),
\quad \lambda_2 \to 0.
\end{split}
\end{equation}

We can now substitute equations~\eqref{eq:limpi}
and~\eqref{eq:limitAbel} into the theta functions in (\ref{eq:para})
to obtain their orders of magnitude as $v_2\rightarrow 0$. Let
$\mathcal{A}$ be a constant vector that is independent of $y$ and
$v_2$. We have
\begin{equation}
\theta(s) = \theta\left(u(y)+\mathcal{A}\right)=\sum_{m\in\mathbb{Z}}e^{i\pi
\Pi m^2+2i\pi \left(u(y)+\mathcal{A}\right)m}.
\end{equation}
The arguments of the exponentials become
\begin{equation}
\begin{split}
m^2\Pi+2\left(u(x)+\mathcal{A}\right)m&=
m(m+1)\Pi+2\left(\frac{1}{\pi i}\left(\frac{\delta}{2\lambda_2}\right)^{\frac{1}{2}}+\mathcal{A}\right)m\\
&\quad +O\left(\left(\frac{\delta}{\lambda_2}\right)^{\frac{3}{2}}\right)+O(\lambda_2),
\quad \lambda_2 \to 0.
\end{split}
\end{equation}
The asymptotic behaviour of the period $\Pi$ in
(\ref{eq:limpi}) gives
\begin{equation}\label{eq:thetalim}
\theta\left(u(y)+\mathcal{A}\right)=1
+e^{-\left(\frac{2\delta}{\lambda_2}\right)^{\frac{1}{2}}-2\pi
i\mathcal{A}}\left(1+O\left(\frac{\delta}{\lambda_2}\right)+O(\lambda_2)\right),
\quad \lambda_2 \to 0.
\end{equation}
By substituting 
\begin{equation}
  \mathcal{A} = -\frac14 \quad \text{and} \quad
  \mathcal{A}=-\frac{N}{2}-\frac{v_1\xi}{2\pi i}-\frac{1}{4} 
\end{equation}
into~\eqref{eq:thetalim}, we see that the matrix elements
in~\eqref{eq:para_1} are bounded by infinity and zero.  Note that,
although the term $v_1\xi /(2\pi i)$ depends on $v_2$ through $\xi$,
since our goal is to study the coefficients of $v_1$, we can always
let $v_1$ be arbitrarily small so that the term $v_1\xi/(2\pi i)$ will
only introduce a negligible error into (\ref{eq:thetalim}). In
particular, we have
\begin{equation}
  \label{eq:theta_order}
  \theta \left(u(y) - \tfrac{N}{2} - \tfrac{v_1\xi}{2\pi i} - \tfrac14
  \right) = O\left(e^{k_1\frac{v_1}{\lambda_2}}\right),
\end{equation}
where $k_1$ is of order $O(1)$ in $v_2$.

From equation~(\ref{eq:omegalim}) and the fact that $\lambda_3=O(1)$
it follows that the term $u(\infty)$ in the constant $H$ in
(\ref{eq:para_2}) remains finite as $v_2 \to 0$.  In fact, we have
\begin{equation}
\label{eq:uinfty}
u(\infty)=\int_{\lambda_3}^{\infty}\frac{ds}{K_0s\sqrt{s^2-\lambda_3^2}}
\Bigl(1+O\left(\lambda_2^2\right)\Bigr)
=\frac{1}{4}+O\left(\lambda_2^2\right), \quad \lambda_2 \to 0.
\end{equation}
Proceeding as in the derivation of (\ref{eq:thetalim}), we see that
also $H$ remains finite and non-zero as $v_2\rightarrow
0$. 

Arguments similar to those ones that led to
equation~\eqref{eq:limitAbel} give 
\begin{equation}
\label{eq:order_F_v1_l2}
\abs{F(y)}=O\left(k_2\frac{v_1}{\lambda_2}\right)\quad \text{and} \quad
\frac{v_1}{y}=O\left(k_3\frac{v_1}{\lambda_2}\right),
\end{equation}
where $y = \pm \lambda_2 + \delta$, and  $k_2$ and $k_3$ are of order
$O(1)$ in $v_2$.  Let us write 
\begin{equation}
  \label{eq:constan_k}
  k := \max \left\{k_1,k_2,k_3 \right \}.
\end{equation}
Equation~\eqref{eq:order_out_param} follows by combining
formulae~\eqref{eq:gammaasym}, \eqref{eq:theta_order}
and~\eqref{eq:uinfty}, and equation~\eqref{eq:order_exp} is
a simple consequence of\eqref{eq:order_F_v1_l2}.
\end{proof}



We are now ready to prove
\begin{corollary}
\label{cor:order}
Let $0< \epsilon <3 $. Then,
formulae~\eqref{eq:diffidasym} hold for $v_2 > N^{-3
    +\epsilon}$ with an error terms of order
\begin{equation}
  \label{eq:error_diffid}
  \mathcal{E} = O\left(N^{-\frac{\epsilon}{9}}e^{k\frac{v_1}{\lambda_2}}\right).
\end{equation}
\end{corollary}
\begin{proof}
Combining equation~\eqref{eq:error_term_tc} with the asymptotic
formulae~\eqref{eq:as_for_v2} gives
\begin{equation}
\mathcal{E}=O\left(\frac{v_2^{\frac{1}{3}}
e^{k\frac{v_1}{\lambda_2}}}{N\delta^{2}}\right).
\end{equation}
Then, equation~\eqref{eq:error_diffid} follows by setting
$v_2=O(N^{-3+\epsilon})$ and
\begin{equation}
   \delta=O\left(N^{-1+\frac{2\epsilon}{9}}\right)
   =O\left(\lambda_2N^{-\frac{\epsilon}{9}}\right).
\end{equation}
\end{proof}
\section{Asymptotics of the ensemble average $E_N(z,t)$}
\label{sec:ensemble}
We are now in a position to give an asymptotic formula for the
ensemble average~\eqref{eq:main} and complete the proof of
Theorem~\ref{thm:main}.

Let us translate formulae~\eqref{eq:diffidasym} back into the original
variables $z = N \sqrt v_2$ and 
\newline
$t = \sqrt{N}v_1$. We obtain
\begin{subequations}
\begin{align}
\frac{\p \log G_N}{\p
z}&=z\left(\frac{1}{2}-\frac{v_2}{16}
\left(\left(\frac{1}{\lambda_2^2}-\frac{1}{\lambda_3^2}\right)^2
+\frac{8}{\lambda_1^4}\right)\right)+O(zN^{-1-\frac{\epsilon}{9}}),\\
\frac{\p \log G_N}{\p
t}&= \frac{\pi i\lambda_1^2}{v_2K_0\xi } \frac{\p}{\p
t}\log\biggl(\theta\left(u(\infty)+\varsigma-\tfrac{t\xi}{2\pi
i\sqrt{N}}\right)\theta\left(u(\infty)+\varsigma+\tfrac{t\xi}{2\pi
i\sqrt{N}}\right)\biggr) \nonumber \\
&\quad -\frac{t}{2N\lambda_1^2}
-\frac{C}{2\lambda_2\lambda_3\sqrt{N}} 
+O\left(N^{-\frac{1}{2}-\frac{\epsilon}{9}}\right),
\end{align}
\end{subequations}
where the $\lambda_j$'s, $K_0$ and $\xi$ are functions of $z$
only. Proposition~\ref{pro:order} and Corollary~\ref{cor:order} imply
that these formulae hold only in the range
$c_1 N^{-\frac{1}{2}}<z< c_2N$, where $c_1$ and $c_2$ are constants
independent of $z$ and $N$. 

Equation~(\ref{eq:C}) implies that
\begin{equation}
C=\frac{v_1\lambda_2\lambda_3}{2K_0\pi}+O\left(\lambda_2^2\right), \quad
\lambda_2 \to 0.
\end{equation}
From this expression, the definition of $\xi$~\eqref{eq:xi} and the
behaviour of the $\lambda_j$'s as $v_2 \to 0$ (see
equation~\eqref{eq:order_lambda}), we obtain
\begin{subequations}
\begin{align}
\frac{\p \log G_N}{\p
z}&=\frac{z}{2}-\frac{3}{2^{\frac{4}{3}}}N^{\frac{2}{3}}z^{\frac{1}{3}}
+O\left(\frac{z^{\frac{5}{3}}}{N^{\frac{2}{3}}}\right)
+O\left(zN^{-1-\frac{\epsilon}{9}}\right),\\
\frac{\p \log G_N}{\p
t}&= -\frac{1}{2}\frac{\p}{\p
t}\log\biggl(\theta\left(u(\infty)+\varsigma-\tfrac{t\xi}{2\pi
i\sqrt{N}}\right)\theta\left(u(\infty)+\varsigma+\tfrac{t\xi}{2\pi
i\sqrt{N}}\right)\biggr) \nonumber \\
& \quad + \left(\frac{tN^{\frac{1}{3}}}{2^{\frac{2}{3}}z^{\frac{4}{3}}}-\frac{t}{16N}\right)\left(1+O\left(v_2^{\frac{2}{3}}\right)\right)
+O\left(N^{-1-\frac{\epsilon}{9}}e^{k\frac{v_1}{\lambda_2}}\right).
\end{align}
\end{subequations}
Integrating these formulae gives
\begin{equation}
\label{eq:Bapprox}
\begin{split}
E_{N}(z,t)&=B_{N}^{\epsilon}\exp\left(\frac{z^2}{4}-\frac{9}{2^{\frac{10}{3}}}\left(N^{\frac{2}{3}}z^{\frac{4}{3}}-N^{\frac{2\epsilon}{3}}\right)+
\frac{t^2N^{\frac{1}{3}}}{2^{\frac{5}{3}}z^{\frac{4}{3}}}\right)
\\
&\quad \times\biggl(\theta\left(u(\infty)+\varsigma-\tfrac{t\xi}{2\pi
i\sqrt{N}}\right)\theta\left(u(\infty)+\varsigma+\tfrac{t\xi}{2\pi
i\sqrt{N}}\right)\biggr)^{\frac{1}{2}}\\
&\quad \times\left(1+O\left(\frac{z^{\frac{8}{3}}}{N^{\frac{2}{3}}}\right)
+O\left(N^{-\frac{1}{2}+\frac{\epsilon}{9}}
e^{k\frac{v_1}{\lambda_2}}\right)+O\left(\frac{t^2}{N}\right)\right),
\end{split}
\end{equation}
 where $B_{N}^{\epsilon}$ is the constant
\begin{equation}
B_{N}^{\epsilon}= \int_{\mathbb{R}^N}\left(\prod_{j=1}^N 
e^{-\frac{1}{2N^{-\epsilon}x_j^2}}\right) P_{\mathrm{GUE}}(x_1,\dotsc,x_N) d^Nx,
\end{equation}
which is independent of $t$.  

The error term $O\left(z^{\frac83}/N^{\frac23}\right)$ in
equation~\eqref{eq:Bapprox} diverges unless 
$z=O\left(N^{1/4}\right)$.  Therefore, we will restrict the validity
of ~\eqref{eq:Bapprox} to $c_1N^{-\frac12}<z < c_2N^{\frac14}$.

We now use the expression of the period $\Pi$ in equation~(\ref{eq:limpi})
to simplify the theta functions in formula~\eqref{eq:Bapprox}.  Combining
the definition of the theta function (\ref{eq:thetadef}) and the
limit of $u(\infty)$ as $\lambda_2 \to 0$ (\ref{eq:uinfty}) gives
\begin{equation}
\theta\left(u(\infty)+\varsigma\pm\tfrac{t\xi}{2\pi
i\sqrt{N}}\right)= 1+\sum_{m\neq
0}\left(\frac{\lambda_2}{16\lambda_3^2}e^{c_N\lambda_2}\right)^{m^2}
e^{2\pi im\left(\tfrac{1}{4}+\varsigma\pm\tfrac{t\xi}{2\pi
i\sqrt{N}}+O(\lambda_2^2)\right)}.
\end{equation}
where $c_N$ is a constant of order $O(1)$ in $\lambda_2$.
Hence, we have
\begin{equation}
\theta\left(u(\infty)+\varsigma\pm\tfrac{t\xi}{2\pi
i\sqrt{N}}\right)=1+\left(\frac{\lambda_2}{32}e^{c_N\lambda_2}\right)
\cos\biggl(2\pi\left(\tfrac{1}{4}+\varsigma\pm\tfrac{t\xi}{2\pi
i\sqrt{N}}\right)\biggr)+O\left(\lambda_2^3\right).
\end{equation}
Substituting this formula back into (\ref{eq:Bapprox}) and letting
$N\rightarrow\infty$ followed by $\epsilon\rightarrow 0$ gives
\begin{equation}
\begin{split}
\lim_{\epsilon \to 0}
\lim_{N\rightarrow\infty}\frac{E_{N}(z,t)}{B^{\epsilon}_{N}}
\exp\left(\frac{9}{2^{\frac{10}{3}}}
\left(N^{\frac{2}{3}}z^{\frac{4}{3}}-N^{\frac{2\epsilon}{3}}\right)-\frac{t^2N^{\frac{1}{3}}}{2^{\frac{5}{3}}z^{\frac{4}{3}}}\right)&=\exp\left(\frac{z^2}{4}\right).
\end{split}
\end{equation}
This limit completes the proof of Theorem~\ref{thm:main}.

\vspace{.25cm}

\noindent\rule{16.2cm}{.5pt}

\vspace{.25cm}

{\small \noindent {\sl Department of Mathematics \\
                       University of Bristol\\
                       Bristol BS8 1TW, UK  \\
                       Email: {\tt f.mezzadri@bristol.ac.uk}\\
                       Email: {\tt m.mo@bristol.ac.uk}

                       \vspace{.25cm}

                       \noindent  31 March 2009}}

\end{document}